\documentclass[a4paper,twoside]{article}

\usepackage{epsfig}
\usepackage{subcaption}
\usepackage{calc}
\usepackage{amssymb}
\usepackage{amstext}
\usepackage{amsmath}
\usepackage{amsthm}
\usepackage{multicol}
\usepackage{pslatex}
\usepackage{apalike}
\usepackage{algorithm2e}
\usepackage[bottom]{footmisc}
\usepackage{tikz}
\usetikzlibrary{arrows.meta, positioning, calc}
\usepackage{tabularx}
\usepackage{booktabs}
\usepackage{SCITEPRESS}     

\begin{document}


\title{A didactical-driven teacher assistant for a dimensional modeling course}

\author{\authorname{Laurent Brisson\sup{1}\orcidAuthor{0000-0002-5309-2688}, Maria Teresa Segarra\sup{1}\orcidAuthor{0000-0003-2714-842X} and Grégory Smits\sup{2}\orcidAuthor{0000-0002-0436-9273}}
\affiliation{\sup{1}{IMT Atlantique, Lab-STICC, UMR CNRS 6285, Brest, France}}
\email{\{laurent.brisson,mt.segarra,gregory.smits\}@imt-atlantique.fr}
}


\abstract{%
Educational chatbots powered by large language models (LLMs) show promising effects on learning outcomes, yet most systems delegate pedagogical decisions such as content selection and didactic structuring implicitly to the LLM, making tutoring strategies difficult to trace, evaluate, and reproduce. This paper presents a didactical-driven teacher assistant for a French-language university course on dimensional modelling, operating without commercial LLM budget or GPU infrastructure. The architecture formalises the instructor's pedagogical reasoning into deterministic modules that handle intent detection, concept linking, and didactic approach selection before any text is generated; the LLM acts solely as a linguistic executor. Evaluation on 195 authentic student questions addresses two research questions. First, we show that standard semantic retrieval alone does not reliably recover the pedagogically required content, thereby justifying the upstream orchestration strategy adopted in our architecture (RQ1). Second, compared to free-tier LLMs whose detection performance varies widely across models and which produce errors silently, the deterministic pipeline achieves high pair precision (73\%) with full traceability and explicit abstention, though its limited coverage confirms that the detection strategy requires further refinement (RQ2).
}


\onecolumn \maketitle \normalsize \setcounter{footnote}{0} \vfill

\maketitle


\section{Introduction}\label{sec:intro}

Educational chatbots have emerged as a prominent application of artificial intelligence in higher education, attracting substantial research attention in recent years. Systematic reviews report positive effects across multiple learning outcomes, including academic achievement, motivation, engagement and self-efficacy~\cite{Debets2025,Elkot2025}. Beyond content delivery, LLM-powered chatbots open new possibilities for formative assessment: recent work demonstrates that dialogue-level knowledge tracing can outperform traditional methods in predicting student response correctness, enabling finer-grained diagnosis of misconceptions during open-ended tutoring interactions~\cite{Scarlatos2025}.

To ground chatbot responses in verified curricular content, several systems combine LLMs with Retrieval-Augmented Generation (RAG), reporting substantial accuracy gains over standalone LLM baselines~\cite{Liu2025}. Other approaches structure retrieval around explicit knowledge representations: domain ontologies and rule-based reasoning can drive pedagogical progression before any content is generated~\cite{Hafner2024,Abdelmagied2025}. More broadly, separating an explicit content planner from the generation step has been shown to improve the pedagogical relevance of generated responses~\cite{Shridhar2022}.

Despite these advances, three structural limitations persist. First, many educational chatbots lack a solid theoretical foundation: they are deployed without grounding in established pedagogical or psychological frameworks~\cite{Debets2025,Elkot2025}. Second, LLMs used as tutors exhibit error rates that remain too high for autonomous instructional roles, with substantial performance variability across knowledge domains~\cite{May2024,Maurya2025}. Third, even in RAG-augmented architectures, the LLM retains an implicit decision-making role in selecting and organising pedagogical content, making tutoring strategies difficult to trace, evaluate, and reproduce independently of the underlying model~\cite{Liu2025}. Yet, no existing system fully externalises the chain of pedagogical decisions into independently evaluable deterministic modules.

This work responds to a concrete pedagogical need: providing students of a French-language university course on dimensional modelling with a chatbot capable of answering their conceptual questions while faithfully reflecting the instructor's teaching practice. The starting point is a formalisation of the instructor's pedagogical reasoning: when answering a student question, an experienced instructor implicitly identifies the student's intent, determines which concepts are involved, mobilises related concepts needed for a coherent response, and selects an appropriate didactic strategy. This reasoning process has been systematised into a deterministic orchestration architecture (Section~\ref{sec:architecture}), and the course material has been jointly restructured to serve both human reading and machine processing, in alignment with a knowledge graph of 43~domain concepts.

This work operates under constraints representative of many real-world deployments but rarely studied: no budget for commercial LLMs and no GPU infrastructure. These constraints are largely absent from the educational chatbot literature, which predominantly evaluates English-language systems backed by commercial models~\cite{Wang2023,Letourneau2025,Debets2025}. Specifically, we address two research questions. First, we examine whether semantic retrieval alone is sufficient to retrieve the pedagogically required content within a realistic prompt budget in our domain-specific setting~(RQ1). Second, we evaluate to what extent a deterministic rule-based orchestration pipeline can achieve reliable pedagogical routing (covering intent detection, concept identification, and end-to-end alignment) when compared to free-tier LLMs in a French, domain-constrained context~(RQ2).

The contributions of this paper are threefold: (i)~an architecture grounded in a formalisation of the instructor's pedagogical reasoning, in which each identified intent maps to a didactic approach and deterministic modules handle the full orchestration chain while the LLM acts solely as a linguistic executor; (ii)~a modular evaluation protocol that assesses each decision stage independently, following the component-level approach advocated by~\cite{Maurya2025}; and (iii)~empirical results on both a controlled synthetic corpus and a real corpus of 195~questions collected during production use by 24~students. Section~\ref{sec:related} reviews related work. Section~\ref{sec:architecture} presents the system architecture. Section~\ref{sec:tech-eval} details the technical evaluation. Section~\ref{sec:discussion} discusses limitations and future directions, and Section~\ref{sec:conclusion} concludes.


\section{Related work}\label{sec:related}

This section discusses the proposed approach in relation to the current landscape of AI-based teaching assistants and existing initiatives for controlling LLM-based components. The former topic aligns with the functional objective we target and the latter guides the conception of the developed teacher assistant.

\subsection{AI-based Teaching Assistants}

A direct use of foundation models in education is to consider them as teacher proxy to answer learners' questions in remote classes or when the teacher is not available~\cite{hicke2023ai,alsafari2024towards}. Experiments confirm the ability of such models to generate appropriate answers to a huge number of learners’ questions that proves the usefulness of such artificial assistants to help real teachers manage remote and hybrid classes or very large groups of students. The teaching assistant presented in~\cite{liu2024hita} goes a step further, as it extends the foundation model, using Retrieval-Augmented Generation (RAG, a technique that augments LLM generation with retrieved documents) techniques, to rely on pedagogical resources provided by the teacher. RAG techniques to specialize foundation models can go really beyond a prompt contextual expansion with the plain text of pedagogical resources. Knowledge graphs (KG) play crucial roles within RAG strategies to incorporate a semantic modelling of the application context that cannot be directly extracted from the queried resources~\cite{baek2023knowledge}. A KG is for instance used in~\cite{bui2024cross} to link different heterogeneous data sources so as to improve the expressiveness of their educational query-answering chatbot. In contrast to these approaches, which rely on rich or cross-source knowledge structures, the present work adopts a lightweight, instructor-curated KG that is sufficient to drive deterministic pedagogical orchestration without requiring external knowledge bases. In this landscape, it is crucial to control every task entrusted to RAG-based teaching assistants and to maintain human oversight so that automatic assistance aligns with pedagogical concerns.

\subsection{Evaluation of LLM-based educational systems}

Current evaluations of LLM-based educational systems predominantly assess end-to-end output quality, such as fluency, factual accuracy, or user perceptions, rather than the conformity of pedagogical decisions to explicit, verifiable principles~\cite{Maurya2025}. Systematic reviews confirm this pattern: most educational chatbots are evaluated through perceptual measures and short-term designs, with few studies examining whether the system's instructional behaviour aligns with its intended pedagogical logic~\cite{Debets2025,Elkot2025}. Moreover, LLM-education benchmarks and pedagogy-driven evaluation frameworks remain largely disconnected, the former focusing on task performance, the latter on didactic principles, with few attempts to bridge the two systematically~\cite{Wang2023}. This makes it difficult to determine whether an observed failure originates from content retrieval, pedagogical routing, or text generation. A decomposed architecture in which each decision stage (intent detection, concept identification, didactic structuring) is an independently evaluable module directly addresses this limitation, enabling component-level diagnosis following the approach advocated by~\cite{Maurya2025}.


\section{System architecture}\label{sec:architecture}


The proposed system is a didactical-driven teacher assistant that supports students during a dimensional modeling course. Its architecture formalises the reasoning an instructor follows when answering a student question: identify the pedagogical intent, determine which domain concepts are involved, mobilise related concepts through a knowledge graph, and construct a response structured by the didactic approach associated with that intent (see Table~\ref{tab:approaches}). The assistant grounds every answer in instructor-authored course material and systematically provides \emph{permalinks} to the relevant teaching pages. Two design goals further constrain the implementation: (i)~keep operational costs low and (ii)~avoid unconstrained LLM decision-making.


\subsection{Overview and design rationale}\label{sec:arch-overview}

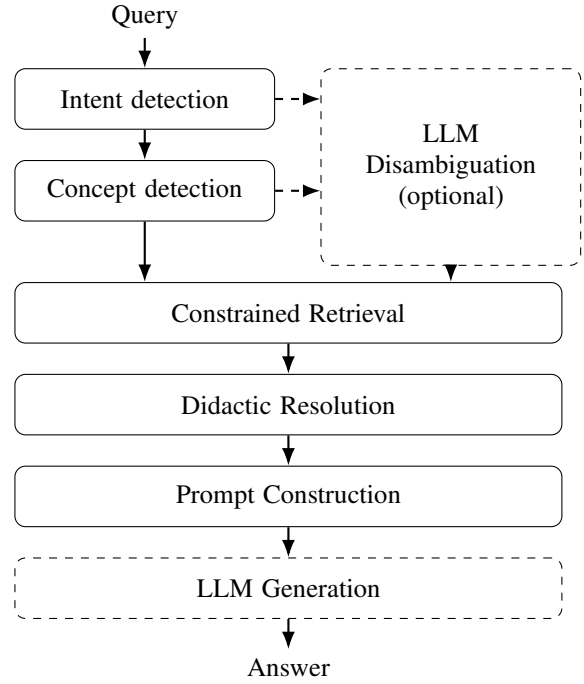
\begin{figure}[t]
\centering
\begin{tikzpicture}[
  node distance=4mm and 6mm,
  full/.style={draw, rounded corners, align=center, text width=7cm, minimum height=8mm, inner ysep=2mm},
  half/.style={draw, rounded corners, align=center, text width=3.2cm, minimum height=8mm, inner ysep=2mm},
  neural/.style={draw, rounded corners, dashed, align=center, text width=3.2cm, minimum height=26mm, inner ysep=2mm},
  neuralfull/.style={draw, rounded corners, dashed, align=center, text width=7cm, minimum height=8mm, inner ysep=2mm},
  arrow/.style={-Latex, thick},
  darrow/.style={-Latex, thick, dashed}
]

\node (q) {Query};

\node[half, below=of q] (id) {Intent detection };
\node[half, below=of id] (cl) {Concept detection};

\node[neural, anchor=north west] (dis) at ($(id.north east) + (0.6, 0)$) {LLM\\Disambiguation\\(optional)};

\node[full, below=8mm of cl, xshift=1.9cm] (cr) {Constrained Retrieval};
\node[full, below=of cr] (dr) {Didactic Resolution};
\node[full, below=of dr] (pc) {Prompt Construction};
\node[neuralfull, below=of pc] (gen) {LLM Generation};

\node[below=of gen] (a) {Answer};

\draw[arrow] (q) -- (id);
\draw[arrow] (id) -- (cl);

\draw[darrow] (id.east) -- (id.east -| dis.west);
\draw[darrow] (cl.east) -- (cl.east -| dis.west);

\draw[arrow] (cl.south) -- (cl.south |- cr.north);
\draw[darrow] (dis.south) -- (dis.south |- cr.north);

\draw[arrow] (cr) -- (dr);
\draw[arrow] (dr) -- (pc);
\draw[arrow] (pc) -- (gen);
\draw[arrow] (gen) -- (a);

\end{tikzpicture}
\caption{Architecture overview. Dashed boxes indicate LLM-assisted stages (Disambiguation and Generation), while solid boxes represent deterministic rule-based processes.}
\label{fig:pipeline}
\end{figure}

A central principle of the system is the separation of pedagogical orchestration from text generation. Instructional decisions (intent recognition, concept selection, chunk typing, and prompt structure) are handled by deterministic modules whenever possible; LLM calls are restricted to two narrowly scoped operations: (i) controlled disambiguation when deterministic signals are insufficient, and (ii) surface-form generation from pre-selected content and pre-established structural instructions. The orchestration layer determines \emph{what} content to present, \emph{in what order}, and \emph{at what level of abstraction}; the LLM determines only \emph{how} to formulate the response in natural language given these constraints. This reduces the impact of the chosen LLM on pedagogical behaviour, which is important because the assistant is designed to operate with lightweight, free-to-use models (e.g., \texttt{gemini-2.0-flash}, \texttt{mistral-small-3.1-24b}, or an \texttt{openrouter-free} proxy). This separation ensures pedagogical traceability (every decision maps to an explicit rule), enforces correctness through deterministic unit-testable logic, and makes the pipeline model-agnostic.
The processing pipeline proceeds through six stages: (i) intent detection, (ii) concept detection, (iii) constrained retrieval, (iv) didactic resolution, (v) prompt construction, and (vi) LLM generation, as illustrated in Figure~\ref{fig:pipeline}.

\subsection{Query analysis}\label{sec:query-analysis}

The first stage determines two properties of the student's question: the \emph{pedagogical intent} (what type of didactic response is requested) and the \emph{domain concepts} (which entities from the course knowledge graph are referenced).

\subsubsection{Intent detection}\label{sec:intent-detection}

Five pedagogical intents are defined, forming an exhaustive set grounded in the instructor's experience designing and teaching the course: \textsc{define}, \textsc{explain}, \textsc{illustrate}, \textsc{compare}, and \textsc{identify}. \textsc{Identify} is the inverse of \textsc{define}: the student provides a description or paraphrase and expects the assistant to name the corresponding concept.
Intent detection is implemented as a deterministic lexicon-based classifier. The question is tokenised using a French-language NLP model (spaCy \texttt{fr\_core\_news\_sm}), and tokens are matched against four categories of linguistic signals: action verbs (weight $+3$), logical connectors (weight $+2$), discourse markers (weight $+4$), and syntactic patterns (weight $+5$). For each candidate intent, the detector computes a weighted sum of matching signals and selects the highest-scoring intent if it reaches a minimum confidence threshold of~3. 
If no intent reaches the threshold, the question is labelled \textsc{uncertain}. In addition, when multiple intents reach the threshold simultaneously, the system does not commit to a deterministic decision because the subsequent pattern-based analysis cannot reliably associate each intent to a specific concept. In both cases (\textsc{uncertain} or multi-intent), the pipeline triggers a controlled LLM-based disambiguation step to resolve the ambiguity among competing intent--concept pairs and to produce a coherent set of associations for downstream processing.

\subsubsection{Concept detection}\label{sec:concept-detection}

The second component links the question to concepts in the course knowledge graph. It generates contiguous n-grams (up to the maximum label length) from the tokenised question, normalises them (lowercasing, accent stripping), and attempts exact matching against normalised concept labels. When exact matching fails, it applies a lightweight fuzzy strategy that combines (i) token overlap to handle reordering and multiword expressions, (ii) directional containment to handle abbreviated mentions, and (iii) character-level similarity to absorb typographical noise.
When concept detection yields no concept, or yields an incompatible outcome for the expected processing (e.g., too many candidates), the system delegates concept identification to an LLM-based fallback constrained to return valid concept identifiers from the knowledge graph. This fallback is narrow in scope and preserves the deterministic intent outcome whenever it is confidently detected. In the rare case of a tie between two concept candidates, the system retains the concept whose match contains the highest number of significant tokens.

\begin{table*}[h]
\caption{Didactic approaches. Each approach targets a Bloom level and
defines core chunk types that structure the generated response, as well
as its ontology-enrichment strategy.
{\footnotesize $^\dag$~Required chunk type (retrieval failure triggers
graceful degradation). Other listed types use a fallback strategy:
retrieved when available for the target concept.}}
\label{tab:approaches}
\centering\small
\begin{tabularx}{\textwidth}{@{}ll l X X@{}}
\toprule
\textbf{Approach} &
\textbf{Bloom} &
\textbf{Concept scope} &
\textbf{Core chunk types} &
\textbf{Ontology enrichment} \\
\midrule
Definition
  & 1--2
  & Single concept
  & \texttt{definition}$^\dag$, \texttt{notation}, \texttt{rule}, \texttt{warning}
  & Always on definition (depth 1) \newline
    Fallback on notation \\
\addlinespace
Identification
  & 1--2
  & Single concept
  & \texttt{definition}, \texttt{illustration}, \texttt{explanation}
  & Always on definition (depth 1) \newline
    Fallback on illustration \\
\addlinespace
Illustration
  & 2--3
  & Single concept
  & \texttt{illustration}$^\dag$, \texttt{definition}
  & Fallback on illustration \newline
    Fallback on counter-example \\
\addlinespace
Explanation
  & 2--4
  & Single concept
  & \texttt{explanation}$^\dag$, \texttt{process}, \texttt{rule}, \texttt{definition}
  & Always on explanation (depth 1) \newline
    Fallback on definition \\
\addlinespace
Comparison
  & 4--5
  & Two or more
  & \texttt{comparison}$^\dag$, \texttt{definition}
  & Always on comparison \newline
    Fallback on definition \\
\addlinespace
Complex synthesis
  & 4--6
  & Multi-concept
  & \texttt{definition}$^\dag$, \texttt{explanation}, \texttt{comparison}, \texttt{illustration}
  & Always on definition + comparison \newline
    Fallback on explanation \\
\bottomrule
\end{tabularx}
\end{table*}

\subsection{Knowledge base and constrained retrieval}\label{sec:knowledge-retrieval}

\subsubsection{Knowledge graph}\label{sec:ontology}

Dimensional modelling is the design of decision-support databases whose structure is guided by the decision-maker's analytic needs and by the application of modelling patterns to business processes. The course knowledge is represented as a lightweight \emph{knowledge graph} (KG) containing 43 concepts organised into thematic groups. Each concept is connected through two families of relations. \emph{Structural relations} (\texttt{is\_a}, \texttt{part\_of}) encode taxonomy and composition. \emph{Didactic relations} (\texttt{precedes}, \texttt{confused\_with}, \texttt{mutually\_exclusive\_with}) encode prerequisite ordering, commonly confused pairs, and incompatible alternatives. These relations are curated by the instructor and support controlled enrichment (Section~\ref{sec:retrieval}) and deterministic follow-up suggestions.

\subsubsection{Editorial corpus}\label{sec:grammar}

The course material is authored as structured Markdown documents for a static website and was intentionally revised by the instructor for dual readability: human and machine. Each concept occupies a dedicated page (except for closely related concepts that must be presented together) and page-level metadata identify the associated concept among the 43~concepts of the knowledge graph. Within each page, content is segmented into discrete chunks following the document's declared structure: chunk boundaries and sizes are determined by the markup itself rather than by token count or sliding windows. Nine chunk types are defined: \texttt{definition}, \texttt{notation}, \texttt{rule}, \texttt{warning}, \texttt{explanation}, \texttt{illustration}, \texttt{special\_case}, \texttt{comparison}, and \texttt{shared\_context}. Both the chunk type and concept identifier are read directly from section headings. Each chunk is annotated with this metadata and indexed in a ChromaDB store where these fields are preserved as filterable attributes.

\subsubsection{Constrained retrieval}\label{sec:retrieval}

Constrained retrieval ensures that content selection is governed by an instructor-defined pedagogical role rather than by unconstrained semantic similarity. Once the intent and concept(s) have been identified, the system selects a didactic approach and queries the store with metadata filters on \texttt{chunk\_type} and \texttt{concept\_id}, optionally extending the concept set via one-step KG traversal when the approach specifies ontological enrichment. The vector store acts as a \emph{typed metadata index}: canonical chunks (definitions, notations, rules) are retrieved by type and concept identifier, not by similarity to the student's question, specifically to avoid confusing short canonical chunks with longer semantically rich passages. If constrained retrieval returns no chunk for a required type, the system returns a static reformulation message that (i) restates the assistant's scope and (ii) invites the student to rephrase, rather than silently switching to unconstrained retrieval.


\subsection{Didactic resolution and generation}\label{sec:resolution-generation}

\subsubsection{Didactic approaches}\label{sec:approaches}

We elicited and transcribed the instructor's actual response practice for each pedagogical intent into a small set of declarative approaches. These approaches, summarised in Table~\ref{tab:approaches}, are specified in configuration files that define, for each intent and target Bloom level, (i)~which chunk types to retrieve, (ii)~whether to enrich the concept set using the knowledge graph, and (iii)~how to structure the prompt sections provided to the LLM.

The knowledge graph may enrich the concept set with structurally related concepts (parents, children, or contrasted concepts) at bounded depth. Ontology enrichment is controlled per chunk type.
Consider the query \textit{"Qu'est-ce qu'une clé naturelle~?"} (What is a natural key). The pattern detector identifies \textsc{define}; concept detection matches \emph{clé naturelle} (natural key) to the KG. The \textit{Definition} approach retrieves the definition of \emph{clé naturelle} and, via the \texttt{is\_a} relation, the definition of its parent concept \emph{clé} (key) for taxonomic context. In parallel, the \texttt{confused\_with} relation to \emph{clé métier} (business key) injects a disambiguation note into the prompt, instructing the LLM to distinguish the two concepts.

\subsubsection{Prompt construction and generation}\label{sec:prompt}

The prompt is assembled from (i) a generic system template defining the assistant's role and communication constraints, (ii) approach-specific section instructions, and (iii) the selected chunks, each labelled by type and accompanied by permalinks to the corresponding course pages. The LLM is instructed to produce a response that follows the section order, respects word-count limits, and uses only the provided content. Text generation is performed using lightweight, free cloud models. Because the pedagogical content and structure are determined upstream, replacing the LLM changes mainly fluency and phrasing while keeping pedagogical behaviour under control; the main exception is the constrained disambiguation fallback described in Section~\ref{sec:query-analysis}.


\section{Technical evaluation}\label{sec:tech-eval}



\subsection{Experimental setup}\label{sec:setup}

\cite{Maurya2025} advocate a modular evaluation architecture for generative AI-powered tutoring systems, in which each pedagogical property (guidance strategy selection, factual correctness, contextual relevance) is assessed as a distinct component rather than subsumed into a single end-to-end quality judgement. They observe that current evaluation practices tend to focus on isolated aspects of system performance, and argue that dedicated metrics for each pedagogical dimension are necessary to produce actionable diagnostic insights.

We adopt this principle by evaluating two decision stages of the pipeline independently. Section~\ref{sec:retrieval-eval} examines whether standard semantic retrieval, the content selection mechanism used in most RAG-based educational systems, can reliably recover the chunk types required by each pedagogical intent from the structured course corpus (RQ1). Section~\ref{sec:orchestration-eval} then evaluates the upstream orchestration components, namely intent detection and concept identification, that determine which content to retrieve, comparing deterministic pattern-based detection against three LLM baselines (RQ2). In both cases, the generated text is not assessed: only the correctness of the pedagogical decisions that precede generation.

\paragraph{Evaluation corpora.} Two corpora are used throughout the evaluation. The \emph{synthetic corpus} comprises 96~questions (143~intent occurences, 209~intent-concept pairs) authored by the instructor in French, following a factorial design that crosses three axes: structural cardinality (1~to~3+ concepts), mention quality (canonical, informal, periphrastic), and sentence complexity (short, compound, embedded clause). This corpus provides controlled conditions and ensures coverage of all five pedagogical intents in balanced proportions. The \emph{real corpus} consists of 195~questions (196~intent occurrences, 213~intent-concept pairs) collected over three weeks of production use by 24~students enrolled in the course. Questions were logged with informed consent and annotated post-hoc by the instructor.  Table~\ref{tab:distribution} summarises the intent distribution in each corpus. The two distributions differ substantially: \textsc{explain} dominates real usage (44\% vs 25\% in the synthetic corpus), while \textsc{compare} and \textsc{identify} (which together account for 45\% of the synthetic design) represent only 8\% of real queries. The real corpus also contains 19\% of out-of-scope, social, and follow-up messages that are entirely absent from the synthetic design.

\begin{table}[ht]
\centering
\caption{Intent distribution across evaluation corpora.}
\label{tab:distribution}
\begin{tabular}{lcc}
\toprule
\textbf{Intent} & \textbf{Synthetic} & \textbf{Real} \\
\midrule
DEFINE      & 20\% (29)  & \textbf{27\% (52)} \\
IDENTIFY    & 22\% (31)  & 2\% (4) \\
ILLUSTRATE  & 10\% (14)  & 2\% (4) \\
EXPLAIN     & 25\% (36)  & \textbf{44\% (86)} \\
COMPARE     & 23\% (33)  & 6\% (12) \\
\textit{no valid intent} & 0\%       & \textbf{19\% (38)} \\
\bottomrule
\end{tabular}
\end{table}


\subsection{Retrieval evaluation (RQ1)}\label{sec:retrieval-eval}

The architecture described in Section~\ref{sec:architecture} detects the pedagogical intent and target concept \emph{before} retrieving content. RQ1 asks whether this upstream orchestration is necessary: could a standard RAG pipeline, querying the course corpus directly with the student's question, retrieve the pedagogically appropriate content on its own? To answer this, we evaluate a standalone semantic retrieval system, with no prior intent or concept detection, on both evaluation corpora. For each question, the ground-truth annotation specifies the expected intent and concept(s), which determine, through the approach configuration, the mandatory chunk type required for a correct response (see Table~\ref{tab:approaches}). We measure whether this required chunk type appears among the top-15 results returned when the student's raw question is used as the sole query input.

\paragraph{Metrics.}
Two complementary indicators are reported. The \emph{retrieval rate} is the proportion of queries for which the required chunk appears in the top-15 results. The \emph{conditional Mean Reciprocal Rank} (MRR$|_{\text{found}}$) measures the average rank quality among those queries where the chunk is found:
\begin{equation}\label{eq:mrr}
  \text{MRR}|_{\text{found}}
  = \frac{1}{n_{\text{found}}}
    \sum_{i=1}^{n_{\text{found}}} \frac{1}{\text{rank}_i}
\end{equation}
where $n_{\text{found}}$ is the number of queries for which the required chunk appears in the top-15. A value of~1.0 means the chunk is always at rank~1; $1/15 \approx 0.067$ means it is always at rank~15. This conditional formulation isolates rank quality from recall: a high MRR with a low retrieval rate means ``well-ranked but rarely found.'' Both indicators must be read jointly.  

\paragraph{Method selection.}  
The student's question is encoded as a dense vector (nomic-embed-text-v1.5) and matched against chunk embeddings by cosine similarity in a ChromaDB store; the top-15 results are retained.

\paragraph{Results on the real corpus.} Table~\ref{tab:retrieval-real} reports retrieval performance stratified by intent on the real corpus (188~evaluable pairs out of~213). \textsc{Define} is the most favourable case: the required chunk is found in 89.1\% of queries with an MRR of~0.795, corresponding to a mean rank of approximately~1.3. \textsc{Explain}, the dominant intent in naturalistic use (44\% of real queries), achieves only 58.2\% retrieval with an MRR of~0.444, meaning that when the required chunk is found, it appears around rank~2.3 on average; one query in two returns no relevant explanation chunk at all. \textsc{Compare} fails most severely: only 14.3\% of comparison chunks are retrieved. This failure is structural rather than parametric: a comparison chunk encodes a relationship between two specific concepts, a constraint that embedding similarity cannot capture from a single query vector. \textsc{Illustrate} and \textsc{identify} show respectively 50.0\% and 75.0\% retrieval rates, with only 4~queries each in the real corpus, these observations cannot be considered representative.

\begin{table}[ht]
\centering
\caption{Semantic retrieval on the real corpus (188~evaluable pairs,
         semantic-only variant). MRR$|_{\text{found}}$: conditional
         mean reciprocal rank over queries where the required chunk
         appears in top-15. $^*$~$n{=}4$, insufficient for inference.}
\label{tab:retrieval-real}
\begin{tabular}{lccc}
\toprule
\textbf{Intent} & \textbf{n} & \textbf{Top-15 \%} & \textbf{MRR$|_{\text{found}}$} \\
\midrule
DEFINE      & 55  & 89.1 & 0.795 \\
EXPLAIN     & 91  & 58.2 & 0.444 \\
COMPARE     & 21  & 14.3 & 0.361 \\
ILLUSTRATE$^*$ & 4 & 50.0 & 0.250 \\
IDENTIFY$^*$   & 4 & 75.0 & 0.200 \\
\midrule
\textbf{All} & \textbf{188} & \textbf{62.6} & \textbf{0.599} \\
\bottomrule
\end{tabular}
\end{table}

\paragraph{Comparison with the synthetic corpus.} Table~\ref{tab:retrieval-synth} reports the same metrics on the synthetic corpus (209~evaluable pairs). Global retrieval is higher on the real corpus (62.6\% vs 42.1\%), likely because real queries tend to be longer and frequently contain domain-specific vocabulary from copied exercise prompts or MCQ statements, providing richer signal for embedding matching. However, the hierarchy across intents is stable between corpora: \textsc{define} achieves the highest retrieval rate, \textsc{explain} remains around 50\%, and \textsc{compare} falls to 7.5\%, confirming that the limitations are structural, not corpus-dependent. \textsc{Identify}, with a larger sample (n=45), achieves 68.9\%, which is expected, since concept discovery (matching a student's description to a concept) is precisely the task that embedding similarity is designed for. There is thus a mismatch between retrieval effectiveness and actual usage: the intent for which semantic retrieval performs best (\textsc{identify}, 75\%) accounts for only 2\% of real queries, whereas the dominant intent in practice (\textsc{explain}, 44\% of queries) achieves only 58.2\% retrieval.

\begin{table}[ht]
\centering
\caption{Semantic retrieval on the synthetic corpus (209~evaluable
         pairs, semantic-only variant). IDENTIFY evaluated with
         relaxed criterion (any definitional chunk accepted).}
\label{tab:retrieval-synth}
\begin{tabular}{lccc}
\toprule
\textbf{Intent} & \textbf{n} & \textbf{Top-15 \%} & \textbf{MRR$|_{\text{found}}$} \\
\midrule
DEFINE      & 33  & 72.7 & 0.441 \\
EXPLAIN     & 49  & 49.0 & 0.449 \\
COMPARE     & 67  & 7.5  & 0.195 \\
ILLUSTRATE  & 15  & 26.7 & 0.403 \\
IDENTIFY    & 45  & 68.9 & 0.480 \\
\midrule
\textbf{All} & \textbf{209} & \textbf{42.1} & \textbf{0.441} \\
\bottomrule
\end{tabular}
\end{table}

\paragraph{Implications.} These results show that semantic retrieval does not reliably recover the pedagogically required content, even in the most favourable case (\textsc{define}, 89.1\%). Without prior identification of the pedagogical intent and the target concept, the retrieval system has no means to specify \emph{which type} of chunk it is looking for: it returns content that is semantically close to the question, not content that fulfils the pedagogical role required by the teaching assistance strategy. Constructing a response structured according to a didactic approach (e.g., a definition followed by an explanation and an example) then depends on an implicit, untraceable content selection by the LLM among the retrieved chunks.


\subsection{Intent and concept detection (RQ2)}\label{sec:orchestration-eval}

The retrieval evaluation (Section~\ref{sec:retrieval-eval}) showed that semantic retrieval alone cannot reliably recover the pedagogically required content, particularly for \textsc{explain} and \textsc{compare}. The architecture therefore detects intent and concept \emph{before} retrieval, using a deterministic pattern-based pipeline. RQ2 asks to what extent this rule-based orchestration can achieve reliable pedagogical routing in a domain-specific, French context, and how it compares to \textbf{free-tier} LLMs performing the same task.

\paragraph{Methods.} Four detection methods are compared: (i)~\textbf{pattern}, the deterministic lexicon-based detector proposed in this work, which requires zero LLM calls; (ii)~\textbf{gemini-2.0-flash}, (iii)~\textbf{mistral-small-3.1-24b-instruct}, and (iv)~\textbf{openrouter-free}, a multi-model free proxy with variable backend routing. Each method receives the student's raw question and returns a list of (intent, concept) pairs representing the pedagogical operations required to answer the question.

\paragraph{Metrics.} All detection tasks are evaluated using the F1 score, the harmonic mean of precision (proportion of detected items that are correct) and recall (proportion of expected items that are detected). Intent detection is evaluated by computing F1 per intent, then averaging without weighting by support (F1 macro). This unweighted average ensures that rare intents contribute equally to the aggregate score despite the skewed distribution observed in Table~\ref{tab:distribution}. Concept detection is evaluated by micro-average F1: true positives, false positives, and false negatives are accumulated across all questions before computing a single precision, recall, and F1. Micro-averaging is preferred here because the 43~concepts in the knowledge graph exhibit highly uneven support, and per-concept scores would be dominated by sampling noise from low-frequency concepts.

\subsubsection{Intent detection}\label{sec:intent-results}

Tables~\ref{tab:intent-synthetic} and~\ref{tab:intent-real} report per-intent F1 and F1 macro for intent detection on both corpora. \textsc{Identify} is excluded from the per-intent breakdown and from the F1 macro computation: the real corpus contains only 4~queries with this intent, too few for a meaningful per-intent score, and including it in the unweighted average would let a single classification error shift the aggregate by up to 20~points. On the synthetic corpus (Table~\ref{tab:intent-synthetic}), the pattern detector achieves the highest F1 macro (65.59), with no statistically significant difference from gemini-2.0-flash (60.19, $p{=}0.174$\footnote{$p$ is the $p$-value.}) and a significant advantage over mistral-small (52.67, $p{=}0.003$) and openrouter-free (43.25, $p{<}0.001$). The pattern's advantage is concentrated on \textsc{compare} (F1~71.70 vs 53.33 for all LLMs) and \textsc{illustrate} (F1~66.67 vs 13.33--35.29), intents whose linguistic markers are well captured by the lexicon.

\begin{table}[ht]
\centering
\setlength{\tabcolsep}{2pt}
\caption{Intent detection performance (per-intent F1 and F1 Macro) --- \textbf{Synthetic corpus} ($n=143$~ intent occurrences).\\ $\uparrow$/$\downarrow$: F1 macro
 significantly better/worse than pattern ($p{<}0.05$, bootstrap); absence of symbol
      indicates no significant difference (ns). Synthetic: pattern vs gemini (ns, $p{=}0.174$); vs mistral ($p{=}0.003$); vs openrouter ($p{<}0.001$).}
\label{tab:intent-synthetic}
\begin{tabular}{@{}lcccc@{}}
\toprule
\textbf{Intent} & \textbf{Pattern} & \textbf{Gemini} & \textbf{Mistral} & \textbf{OpenRouter} \\
\midrule
DEFINE      & 80.00          & \textbf{94.55} & 77.14          & 74.51 \\
EXPLAIN     & 44.00          & \textbf{57.58} & 44.90          & 31.82 \\
COMPARE     & \textbf{71.70} & 53.33          & 53.33          & 53.33 \\
ILLUSTRA.  & \textbf{66.67} & 35.29          & 35.29          & 13.33 \\
\midrule
F1 Macro    & \textbf{65.59} & 60.19  & 52.67$\downarrow$ & 43.25$\downarrow$ \\
\textit{No detection}         & 9              & 0              & 0              & 37    \\
\bottomrule
\end{tabular}
\end{table}

\begin{table}[ht]
\centering
\setlength{\tabcolsep}{2pt}
\caption{Intent detection performance (per-intent F1 and F1 Macro) --- \textbf{Real corpus} ($n=196$~intent occurrences).\\ $\uparrow$/$\downarrow$: F1 macro
 significantly better/worse than pattern ($p{<}0.05$, bootstrap); absence of symbol
      indicates no significant difference (ns). pattern vs gemini ($p{<}0.001$); vs mistral (ns, $p{=}0.137$); vs openrouter (ns, $p{=}0.473$)}
\label{tab:intent-real}
\begin{tabular}{@{}lcccc@{}}
\toprule
\textbf{Intent} & \textbf{Pattern} & \textbf{Gemini} & \textbf{Mistral} & \textbf{OpenRouter} \\
\midrule
DEFINE      & 77.27          & \textbf{85.42} & 54.90          & 65.69 \\
EXPLAIN     & 17.02          & \textbf{86.96} & 24.49          & 49.57 \\
COMPARE     & \textbf{66.67 }         & 50.00          & 50.00          & \textbf{66.67} \\
ILLUSTRA.  & 88.89          & \textbf{100.0} & 85.71          & 85.71 \\
\midrule
F1 Macro    & 62.46          & \textbf{80.59}$\uparrow$ & 53.78          & 66.91 \\
\textit{No detection}         & \textbf{133}   & 10             & 0              & 38    \\
\bottomrule
\end{tabular}
\end{table}

On the real corpus (Table~\ref{tab:intent-real}), the ranking changes. Gemini-2.0-flash achieves the highest F1 macro (80.59, $p{<}0.001$ vs pattern), while the pattern reaches~62.46. The pattern's main weakness is \textsc{explain} detection (F1~17.02 on real vs 44.00 on synthetic), which represents 44\% of real queries: the pattern achieves 100\% precision but only 9\% recall, meaning it correctly identifies only prototypical formulations (imperative verb + explicit signal). Neither mistral-small (53.78, $p{=}0.137$) nor openrouter-free (66.91, $p{=}0.473$) differ significantly from the pattern on the real corpus.

\subsubsection{Concept detection}\label{sec:concept-results}  

The course knowledge graph defines 43~concepts that the student may reference in a question. Detecting which concepts are mentioned is challenging because students rarely use the canonical label: they truncate, abbreviate, paraphrase, or use informal variants. To characterise this variability, we partition concepts into three difficulty groups based on the expected detection difficulty. Group~1 contains isolated, unambiguous concepts with a single canonical label and no near-neighbour in the knowledge graph. Group~2 contains multi-formulation concepts where students use truncated forms, orthographic variants, or periphrases instead of the canonical label, e.g., \emph{le grain} for \emph{granularité} (granularity) or \emph{modifications d'une dimension} for \emph{dimension à évolution lente} (slowly changing dimension). Group~3 contains clustered concepts whose labels are lexically adjacent, e.g., \emph{fait additif / semi-additif / non-additif} (additive / semi-additive / non-additive fact), making them prone to partial-match confusion by any method that relies on surface overlap.  

Tables~\ref{tab:concept-synth} and~\ref{tab:concept-real} report concept detection F1 (micro-average) globally and by group. On the synthetic corpus (Table~\ref{tab:concept-synth}), gemini-2.0-flash leads globally (F1~79.15, $p{<}0.001$ vs pattern), followed by mistral-small (73.85, $p{<}0.001$). The pattern (54.44) achieves the highest F1 on Group~1 (84.00), where unambiguous canonical labels are well suited to exact n-gram matching, but drops to~47.21 on Group~3, where lexically adjacent concept labels cause systematic partial-match confusion. This Group~3 degradation affects all methods---even gemini-2.0-flash drops from 90.20 (Group~1) to 78.64 (Group~3)---but it is most severe for the pattern, whose fuzzy matching cannot distinguish labels that share most of their tokens.  

\begin{table}[ht]
\centering
\setlength{\tabcolsep}{2pt}
\caption{Concept detection F1 (micro-average) by difficulty group
         --- \textbf{Synthetic corpus} ($n{=}209$ pairs).
         $\uparrow$: significantly better than pattern
         ($p{<}0.05$, bootstrap).}
\label{tab:concept-synth}
\begin{tabular}{@{}lcccc@{}}
\toprule
\textbf{Group} & \textbf{Pattern} & \textbf{Gemini} & \textbf{Mistral} & \textbf{OpenRouter} \\
\midrule
Group 1   & 84.00 & \textbf{90.20} & 82.14 & 73.68 \\
Group 2   & 59.26 & \textbf{81.16} & 73.97 & 78.43 \\
Group 3   & 47.21 & \textbf{78.64} & 72.55 & 61.62 \\
\midrule
All       & 54.44 & \textbf{79.15}$^\uparrow$ & 73.85$^\uparrow$ & 64.63 \\
\bottomrule
\end{tabular}
\end{table}

On the real corpus (Table~\ref{tab:concept-real}), the pattern improves to~58.62 but remains significantly below gemini-2.0-flash (77.83, $p{<}0.001$) and openrouter-free (70.18, $p{<}0.001$), while not differing from mistral-small (56.06, $p{=}0.742$). The group profile is consistent: the pattern reaches 87.18 on Group~1 but only 46.31 on Group~3, confirming that the difficulty gradient is structural rather than corpus-dependent.

\begin{table}[ht]
\centering
\setlength{\tabcolsep}{2pt}
\caption{Concept detection F1 (micro-average) by difficulty group
         --- \textbf{Real corpus} ($n{=}213$ pairs).
         $\uparrow$: significantly better than pattern
         ($p{<}0.05$, bootstrap).}
\label{tab:concept-real}
\begin{tabular}{@{}lcccc@{}}
\toprule
\textbf{Group} & \textbf{Pattern} & \textbf{Gemini} & \textbf{Mistral} & \textbf{OpenRouter} \\
\midrule
Group 1   & 87.18 & \textbf{90.53} & 82.35 & 84.71 \\
Group 2   & 77.78 & \textbf{93.75} & 67.50 & 85.25 \\
Group 3   & 46.31 & \textbf{72.96} & 48.04 & 65.00 \\
\midrule
All       & 58.62 & \textbf{77.83}$^\uparrow$ & 56.06 & 70.18$^\uparrow$ \\
\bottomrule
\end{tabular}
\end{table}

\subsubsection{Joint orchestration and precision--coverage   trade-off}\label{sec:pair-results}

In a tutoring pipeline, an error on either intent or concept propagates silently through retrieval and prompt construction, producing a pedagogically incorrect response with no recoverable signal for the student. The (intent, concept) pair metric therefore captures the operationally relevant orchestration accuracy: a pair is correct only if both the intent and the concept match the ground-truth annotation.

On the synthetic corpus (Table~\ref{tab:pair-synth}), all three LLM baselines significantly outperform the pattern on pair F1 (bootstrap: gemini-2.0-flash 48.12, $p{<}0.001$; mistral-small 41.54, $p{=}0.001$; openrouter-free 38.15, $p{=}0.037$; vs pattern 27.78). Exact match rates remain low across all methods (11--24\%), reflecting the difficulty of jointly detecting both dimensions on multi-intent questions.

\begin{table}[ht]
\centering
\setlength{\tabcolsep}{2pt}
\caption{Joint (intent, concept) pair detection --- \textbf{Synthetic corpus}
         ($n{=}96$~questions). $\uparrow$/$\downarrow$: significantly
         better/worse than pattern ($p{<}0.05$).}
\label{tab:pair-synth}
\begin{tabular}{@{}lcccc@{}}
\toprule
\textbf{Metric} & \textbf{Pattern} & \textbf{Gemini} & \textbf{Mistral} & \textbf{OpenRouter} \\
\midrule
Pair F1          & 27.78 & \textbf{48.12}$^\uparrow$ & 41.54$^\uparrow$ & 38.15$^\uparrow$ \\
Ex.\ match  & 11\% & \textbf{24\%} & 23\% & 22\% \\
Pair prec.  & 33.11 & 44.67 & 37.60 & \textbf{53.45} \\
\bottomrule
\end{tabular}
\end{table}

On the real corpus (Table~\ref{tab:pair-real}), gemini-2.0-flash dominates (pair F1~67.81, $p{<}0.001$). The pattern (35.19) significantly outperforms mistral-small (24.55, $p{=}0.007$, bootstrap) but is outperformed by openrouter-free (44.50, $p{=}0.015$, bootstrap) and achieves the lowest exact match rate (35.90). However, the pattern shows a radically different precision profile: when it does detect a pair, it is correct 73.21\% of the time, compared to 60.00\% for gemini-2.0-flash, 39.91\% for openrouter-free, and 18.04\% for mistral-small. This asymmetry between low coverage and high precision is the defining characteristic of the deterministic detector.

\begin{table}[ht]
\centering
\setlength{\tabcolsep}{2pt}
\caption{Joint (intent, concept) pair detection --- \textbf{Real corpus}
         ($n{=}195$~questions). $\uparrow$/$\downarrow$: significantly
         better/worse than pattern ($p{<}0.05$).}
\label{tab:pair-real}
\begin{tabular}{@{}lcccc@{}}
\toprule
\textbf{Metric} & \textbf{Pattern} & \textbf{Gemini} & \textbf{Mistral} & \textbf{OpenRouter} \\
\midrule
Pair F1          & 35.19 & \textbf{67.81}$^\uparrow$ & 24.55$^\downarrow$ & 44.50$^\uparrow$ \\
Ex.\ match  & 36\% & \textbf{64\%} & 42\% & 49\% \\
Pair prec.  & \textbf{73.21} & 60.00 & 18.04 & 39.91 \\
\bottomrule
\end{tabular}
\end{table}

Table~\ref{tab:tradeoff} makes this trade-off explicit. The pattern abstains on 68\% of real queries (133 out of 195), but when it commits to a response, it achieves a pair precision of~73.21, meaning that nearly three detected pairs out of four are correct. Gemini-2.0-flash offers the strongest overall profile, with both high coverage (${\sim}95\%$) and acceptable pair precision (60.00). By contrast, mistral-small never abstains yet reaches only 18.04 pair precision: fewer than one detected pair in five is correct.

\begin{table}[ht]
\centering
\setlength{\tabcolsep}{2pt}
\caption{Precision--coverage profile --- \textbf{Real corpus} ($n{=}195$).
         Latency: median over all queries (detection only).
         Pattern abstentions include 37 out-of-scope/social queries
         correctly refused.}
\label{tab:tradeoff}
\begin{tabular}{@{}lcccc@{}}
\toprule
\textbf{Metric} & \textbf{Pattern} & \textbf{Gemini} & \textbf{Mistral} & \textbf{OpenRouter} \\
\midrule
Coverage       & $32\%$ & $95\%$ & $100\%$ & $81\%$ \\
Pair prec.   & \textbf{73.21} & 60.00 & 18.04 & 39.91 \\
Abstention      & \textbf{133} & 10 & 0 & 38 \\
Lat. (ms)  & \textbf{50} & 621 & 4\,287 & 8\,024 \\
\bottomrule
\end{tabular}
\end{table}

The cost function of orchestration errors in a tutoring pipeline is asymmetric: a wrong intent leads to a wrong didactic approach, wrong chunk types, and a pedagogically incorrect response. A refusal to answer, by contrast, triggers a controlled recovery path---typically a reformulation request---that preserves pedagogical integrity. The 37~out-of-scope or social messages correctly refused by the pattern illustrate this safety property. 

These results show that the pattern detector, in its current form, does not provide sufficient coverage to serve as a standalone orchestration method. The combined configuration---pattern first, LLM fallback on abstention---reduces LLM calls by approximately one third but does not improve joint detection accuracy: on the real corpus, the combined pattern+gemini configuration (pair F1~63.50) is significantly lower than gemini-2.0-flash alone (67.81, $p{=}0.005$), indicating that the current pattern introduces errors on queries that the LLM would have handled correctly.

Nevertheless, the evaluation validates the architectural principle that motivated this work. The pattern detector---a zero-cost lexicon baseline---achieves a pair precision of~73.21 on answered queries, demonstrating that deterministic orchestration can reliably resolve a subset of well-formed pedagogical queries with full traceability. Its limitations---particularly the low recall on \textsc{explain}, the dominant intent in practice---are not inherent to the principle of upstream orchestration but to the lexicon-based detection strategy used here as a baseline. The annotated corpus of 195~authentic student queries now supports an incremental improvement trajectory: improving detection coverage through supervised classifiers (e.g., logistic regression, support vector machines) up to transformer-based models (at the cost of GPU infrastructure), and, independently, using selective LLM fallback as an architectural alternative when classifiers abstain. Each step remains independently evaluable using the same modular protocol.


\section{Discussion}\label{sec:discussion}

Throughout this work, we endeavoured to make explicit the chain of pedagogical decisions that an instructor normally performs tacitly, from intent identification to didactic approach selection. This formalisation enables independent evaluation of each decision stage and conditions the transferability of the system, since adapting it to a new course requires an instructor to formalise \emph{their own} pedagogical practice, not merely to configure a tool. The incremental methodology described above, in which a minimal baseline is deployed to bootstrap its own improvement, opens research directions that extend beyond the present system.

\subsection{Toward a transferable bootstrapping methodology}\label{sec:fw-bootstrap}

Most pedagogical orchestration systems reported in the literature presuppose resources that are rarely available in practice: annotated interaction corpora, validated learner models, or large-scale training data. \cite{Francisco2022} further observe that none of the AI-based educational systems surveyed in Computer Science Education made their datasets publicly available. The approach taken here inverts this dependency: the deployed baseline itself produces the annotated resources that subsequent iterations consume, raising a general question: \emph{what is the minimal viable system that can seed its own iterative refinement?} Answering this question, however, goes beyond code sharing. Adapting the system to another course requires constructing a domain ontology, authoring a structured editorial corpus, and calibrating detection lexicons to a new vocabulary and possibly a new language. These steps involve pedagogical expertise that cannot be automated away. Beyond vocabulary, the approach presupposes a course whose content is organizable into clearly delimited concepts and typed pedagogical chunks; courses centered on theorem derivation, algorithmic processes, or code production may require different taxonomies and retrieval strategies. The present architecture is therefore offered as a methodological blueprint rather than a universal solution. A structured \emph{methodological framework} is therefore needed to guide instructors through domain adaptation, accompanied by domain-independent evaluation protocols. Because the architecture evaluates each component independently (intent detection, concept resolution, retrieval, joint orchestration), the same protocol can be applied regardless of domain content, a prerequisite for meaningful cross-course comparison that the field currently lacks. The modular evaluation dimensions proposed by~\cite{Maurya2025} provide a natural foundation for defining such checkpoints across successive iterations.

\subsection{From reactive answering to pedagogical companionship}\label{sec:fw-companion}

The current system is reactive and atomistic: it receives a question, classifies it, retrieves content, and generates a response or abstains. It has no representation of the learner, no awareness of the course timeline, and no capacity to initiate interaction. Moving toward a pedagogical agent that accompanies students throughout a course raises the question of how abstention should be handled. The cost function is asymmetric: a wrong answer propagates silently into a pedagogically incorrect response, while a refusal triggers a controlled recovery. Many abstentions correspond to genuinely underspecified questions, particularly for \textsc{explain}, where the intent is too vague for any system to resolve without clarification. An agent that responds with a well-constructed follow-up question, guiding the student toward a classifiable query, would reframe \textsc{uncertain} as a \emph{pedagogical entry point} rather than a system failure. This is consistent with Socratic principles, where the quality of the question matters as much as the quality of the answer~\cite{Shridhar2022}. Whether a system that helps students formulate better questions outperforms one that always attempts to answer remains an open empirical question.

The knowledge base also conditions what the system can offer. The current corpus is mono-authored: all content originates from the instructor's teaching notes. Integrating heterogeneous sources (reference textbook excerpts, complementary explanatory materials) would broaden coverage and provide students with alternative formulations of the same concepts. Incorporating the course's existing bank of multiple-choice questions would offer a natural resource for revision support, though the integration raises a non-trivial constraint: answers must not be surfaced directly during an interaction, so that the revision value of the question is preserved. In all cases, each new source must be indexed with concept identifiers and chunk types so that constrained retrieval continues to operate on pedagogical role rather than on unconstrained similarity. The retrieval evaluation (Section~\ref{sec:retrieval-eval}) showed that standard semantic similarity struggles to recover canonical content such as definitions and comparisons. Ontology-grounded approaches to retrieval, such as the hypergraph-based method proposed by~\cite{sharma2025} for anchoring RAG in domain ontologies, suggest that enriching concept representations with their structural and didactic neighbourhood could improve retrieval precision for these semantically sparse chunks---a direction that would benefit from the relations already encoded in the knowledge graph.

\subsection{Evaluation as a transversal concern}\label{sec:fw-evaluation}

Each of the directions above requires its own evaluation, and the risk of ad-hoc, non-comparable assessment protocols across educational AI systems is well documented~\cite{Maurya2025}. The modular approach adopted in this paper, where intent detection, concept resolution, and retrieval are assessed independently, extends naturally to future components: a reformulation agent can be evaluated on the classifiability and pedagogical appropriateness of its generated questions; a learner model can be evaluated on the accuracy of its knowledge state predictions. Beyond component-level evaluation, three levels of end-to-end assessment remain necessary: factual and pedagogical quality of the generated responses, perceived usefulness by students, and measurable impact on learning outcomes. Concretely, student-facing impact could be assessed on two complementary axes: performance on conceptual MCQs and accuracy of knowledge mobilisation in engineering projects. Both require longitudinal measurement on a system with stable detection performance and a sufficiently large interaction corpus. The current system and its annotated corpus provide the baseline against which these evaluations can be conducted.


\section{Conclusion}\label{sec:conclusion}
This paper addressed the problem of building a pedagogical teacher assistant for a domain-specific university course under strong operational constraints: no budget for commercial LLMs, no GPU infrastructure, and French instructional language. Rather than delegating pedagogical decisions to a general-purpose language model, we formalised the instructor's own reasoning process (intent identification, concept mobilisation, didactic approach selection) into a knowledge-orchestrated architecture in which deterministic modules handle the full orchestration chain while the LLM acts solely as a linguistic executor on pre-selected, pre-structured content.

The evaluation answered two research questions. Standard semantic retrieval alone cannot reliably recover the pedagogically required content (RQ1): identifying the intent and concept before retrieval is a necessity. A deterministic orchestration pipeline can reliably route well-formed queries with full traceability, though its lexicon-based detection covers only a minority of authentic questions; free-tier LLMs offer broader coverage at the cost of lower precision and no auditability (RQ2). The modular design makes each failure mode independently diagnosable, turning the evaluation into an actionable improvement plan.

This design directly addresses three challenges identified by~\cite{Maurya2025} for GenAI-powered educational systems: hallucination risk is mitigated because the LLM cannot select or suppress content beyond instructor-authored chunks; evaluation is decoupled into pedagogical decision quality and linguistic output quality; and pedagogical behaviour is not tuned through iterative prompt refinement but structured across the entire pipeline, with prompt engineering confined to the final generation step where it governs only writing instructions. This work also illustrates that an incremental methodology, where a minimal system is deployed to bootstrap its own evaluation resources, can produce actionable diagnostics even under severe operational constraints. Generalising this approach, however, remains an open challenge. We believe that a necessary next step for the community is the design of methodological frameworks that guide instructors through the full process, from ontology construction and initial deployment to component-level evaluation, accompanied by domain-independent benchmarking protocols that enable meaningful comparison of systems across courses, institutions, and languages. Beyond component-level assessment, student-facing evaluation covering response quality, perceived usefulness, and measurable learning impact constitutes the next step for this work.

\bibliographystyle{apalike}
{\small
\bibliography{references}}

\end{document}